\documentclass[conference]{IEEEtran}
\IEEEoverridecommandlockouts
\usepackage[T1]{fontenc}
\usepackage{times}
\usepackage{microtype}
\usepackage{algorithmic}
\usepackage{algorithm}
\usepackage{array}
\usepackage{amsmath,amssymb,amsfonts}
\usepackage[caption=false,font=normalsize,labelfont=sf,textfont=sf]{subfig}
\usepackage{cite}
\usepackage{caption}
\usepackage{textcomp}
\usepackage{stfloats}
\usepackage{url}
\usepackage{verbatim}
\usepackage{graphicx}
\usepackage{xcolor}
\usepackage{makecell}
\usepackage{multirow}
\usepackage{amsmath}
\usepackage{fancyhdr}
\usepackage{color}
\usepackage{xcolor} 
\usepackage[marginal]{footmisc}
\usepackage{type1cm}

\usepackage[colorlinks,
            linkcolor=blue,
            anchorcolor=blue,
            citecolor=blue,
            urlcolor=blue
            ]{hyperref}

\begin{document}

\title{Research on a Two-Layer Demand Response Framework for Electric Vehicle Users and Aggregators Based on LLMs\\
}

\author{
\IEEEauthorblockN{Zhaoyi Zhang}
\IEEEauthorblockA{
\textit{Shanghai University of Electric Power} \\
Shanghai, China \\
zhangzhaoyi@mail.shiep.edu.cn}
\and
\IEEEauthorblockN{Chenggang Cui*}
\IEEEauthorblockA{
\textit{Shanghai University of Electric Power} \\
Shanghai, China \\
cgcui@shiep.edu.cn}
\and
\IEEEauthorblockN{Ning Yang}
\IEEEauthorblockA{
\textit{Shanghai University of Electric Power} \\
Shanghai, China \\
nyang@shiep.edu.cn}
\and
\IEEEauthorblockN{Chuanlin Zhang}
\IEEEauthorblockA{
\textit{Shanghai University of Electric Power} \\
Shanghai, China \\
clzhang@shiep.edu.cn}
}

\maketitle
\setcounter{page}{1}
\thispagestyle{plain}
\renewcommand{\thefootnote}{}
\footnote{}
\renewcommand{\footrulewidth}{1pt}
\begin{abstract}
The widespread adoption of electric vehicles (EVs) has increased the importance of demand response in smart grids. This paper proposes a two-layer demand response optimization framework for EV users and aggregators, leveraging large language models (LLMs) to balance electricity supply and demand and optimize energy utilization during EV charging. The upper-layer model, focusing on the aggregator, aims to maximize profits by adjusting retail electricity prices. The lower-layer model targets EV users, using LLMs to simulate charging demands under varying electricity prices and optimize both costs and user comfort. The study employs a multi-threaded LLM decision generator to dynamically analyze user behavior, charging preferences, and psychological factors. The framework utilizes the PSO method to optimize electricity prices, ensuring user needs are met while increasing aggregator profits. Simulation results show that the proposed model improves EV charging efficiency, alleviates peak power loads, and stabilizes smart grid operations.
\end{abstract}

\begin{IEEEkeywords}
Electric Vehicle,Demand Response,LLM,Smart Grid
\end{IEEEkeywords}

\section{Introduction}
As EVs become increasingly prevalent, their large-scale integration into the grid introduces both challenges and opportunities\cite{b1}. The charging behavior of EVs significantly affects the grid due to substantial fluctuations in power demand, which impact both the stability and economic efficiency of the power system\cite{b2}. EV charging demand is typically random and volatile, particularly during peak hours, when demand surges can overload the grid and compromise its stability\cite{b3}. Moreover, uncertainty in EV charging times and volumes poses significant challenges for power system planning and management\cite{b4}.If these charging behaviors are not properly managed and coordinated, they may lead to grid overload, increased operational costs, and potential blackouts\cite{b5}. Therefore, regulating electric vehicle charging behavior through load management schemes is essential for optimizing power system operation and management\cite{b6}.

Currently, electric vehicle charging load management schemes primarily involve several approaches\cite{b7}. Fixed price incentive schemes encourage users to charge during off-peak hours by offering discounts or special rates, thereby alleviating peak load\cite{b8}.  However, due to the absence of dynamic adjustments, these schemes struggle to respond to real-time changes in grid load. Time-of-use pricing strategies incentivize users to charge during low-price periods by varying electricity prices across different time slots\cite{b9}. These strategies are typically implemented in regions with significant load fluctuations, helping balance charging behavior with supply-demand conditions.Real-time pricing strategies adjust electricity prices based on actual supply and demand dynamics in the power market, providing users with immediate charging cost signals. While this approach offers flexibility, price fluctuations could impact the user experience. Smart charging management systems integrate with demand response platforms to dynamically adjust charging power and duration. By utilizing intelligent algorithms and big data analysis\cite{b10}, these systems dynamically optimize charging behavior based on load demand and user habits, offering significant optimization potential. Demand response\cite{b11} mainly encourages users to adjust their charging behavior via price signals and direct control to achieve its objectives.

The emergence of LLMs is transforming the management of EV charging behavior. Using their advanced natural language processing capabilities, LLMs can analyze user charging and discharging habits, behavior patterns, and grid operation data to guide peak shaving and load balancing, thus improving demand response effectiveness.

This paper presents an LLM-based modeling approach for electric vehicle user behavior and integrates it into a bilevel optimization framework to maximize retailer profits. The main
contributions of this paper are listed as follows:

\begin{itemize}
\item Applied LLM to simulate electric vehicle users' charging demand.
\item Proposed a conversational interaction-based method for providing user decision support.
\item Developed a two-layer optimization framework considering user behavior and aggregator profits.
\end{itemize}

\section{Electric Vehicle Demand Response}

In EV charging management, demand response mechanisms are widely used to encourage users to modify their charging behavior via price signals and direct control, aiming to achieve demand response objectives.Price-based demand response utilizes strategies like time-of-use and real-time pricing to incentivize users to charge during low-price periods, thereby distributing the load and mitigating peak demand pressure.This approach is effective for users sensitive to price signals. Direct control-based demand response enables aggregators or power companies to regulate the charging process, reducing power consumption or delaying charging during peak periods. It is primarily implemented for large user groups to achieve precise load control.Incentive-compatible demand response combines price incentives and direct control, encouraging users to charge during discounted periods while compensating them for controlled charging, thereby enhancing participation and user experience.

Although EV demand response mechanisms hold significant potential for balancing electricity supply and demand, their practical implementation faces several challenges. First, the variability in user charging demands introduces randomness in charging time, location, and frequency, complicating load forecasting. Second, frequent fluctuations in electricity prices may degrade user experience, making it challenging for users to adapt to dynamic charging costs.Finally, optimizing demand response to meet grid load requirements remains a challenge for aggregators, requiring further research and model refinement.
\section{Simulation of Electric Vehicle User Charging Behavior}

\subsection{User Behavior Simulation Framework}
To accurately simulate the charging behavior of electric EV users, this study introduces a user behavior simulation framework utilizing a large language model (LLM). This framework is constructed using user behavior data and external environmental factors, featuring a modular design comprising a user profiling module, an environmental state module, and a decision generation module. This framework effectively captures user decision-making processes under dynamic pricing and diverse scenarios.  The detailed framework design is as follows:

\paragraph{User Profiling Module} The user profile module aims to capture the personalized characteristics of EV users, providing data to support the simulation of their behavioral decisions. By detailing individual user attributes and behavioral preferences, this module constructs a comprehensive user profile. The user profile comprises three dimensions: 

\begin{itemize}
\item \textbf{Demographic Characteristics:} Including the user's age, occupation, income level, and place of residence. These characteristics significantly influence the user's charging behavior. For instance, high-income users are more likely to prefer fast charging, while low-income users tend to prioritize charging costs. 
\item \textbf{Psychological Characteristics:} Including the user's environmental awareness, acceptance of new technologies, and risk preferences. Users with greater environmental awareness are more inclined to participate in demand response strategies, while those highly sensitive to technology are more likely to adopt smart charging solutions.
\item \textbf{Charging Preferences: }Including preferred charging times, charging locations, and sensitivity to charging costs. 
\end{itemize}

\paragraph{Environmental State Module} The Environmental State Module characterizes the dynamic charging environment in which users operate, encompassing the current state of the electric vehicle and external conditions.
This module serves as an input for simulating real-time variations in user behavior and primarily encompasses aspects such as battery state, driving demand, and the dynamic electricity price $\lambda(t)$  at time $t$.

\paragraph{Decision Generation Module} The Decision Generation Module serves as the core of the framework, generating user charging decisions based on the user profile and environmental state. User decision logic is simulated using LLM combined with multi-agent modeling to generate behavioral responses under varying electricity prices and external conditions.Its specific functions include:

\begin{itemize}
\item Charging Willingness Assessment: Based on the current state of charge and driving demands, the user decides whether immediate charging is necessary.
\item Charging Time Selection: Taking dynamic electricity prices and time preferences into account, the user chooses the optimal charging period. 
\item Charging Amount Decision: By balancing charging costs and energy demands, the user decides on the exact amount of electricity to be charged. 
\end{itemize}

The decision-generation process is governed by the following formulas: 
\begin{equation}
P_{ri}(t)=f_{agent_i}\left(\lambda(t),SoC,T\right)
\label{eq}
\end{equation}

Where $P_{ri}(t)$ represents the charging power of user $i$ at time $t$; $\lambda(t)$ is the retail electricity price; SoC denotes the battery's state of charge; and $T$ represents external environmental variables (such as temperature). The large language model generates $f_{\text{agent}_i}$ through semantic analysis and reasoning, reflecting the user's sensitivity to dynamic electricity prices and external environmental conditions.

\subsection{LLM-Based User Profiling}
User profiling is a key component in modeling EV users' charging behavior. To effectively capture complex user behavior patterns, this study introduces an LLM-based user profiling approach. Utilizing its advanced semantic analysis and reasoning capabilities, the LLM extracts user features from multidimensional data and dynamically updates profiles, offering crucial support for demand response model optimization.

\begin{figure}[h]
    \centering
    \includegraphics[width=\linewidth]{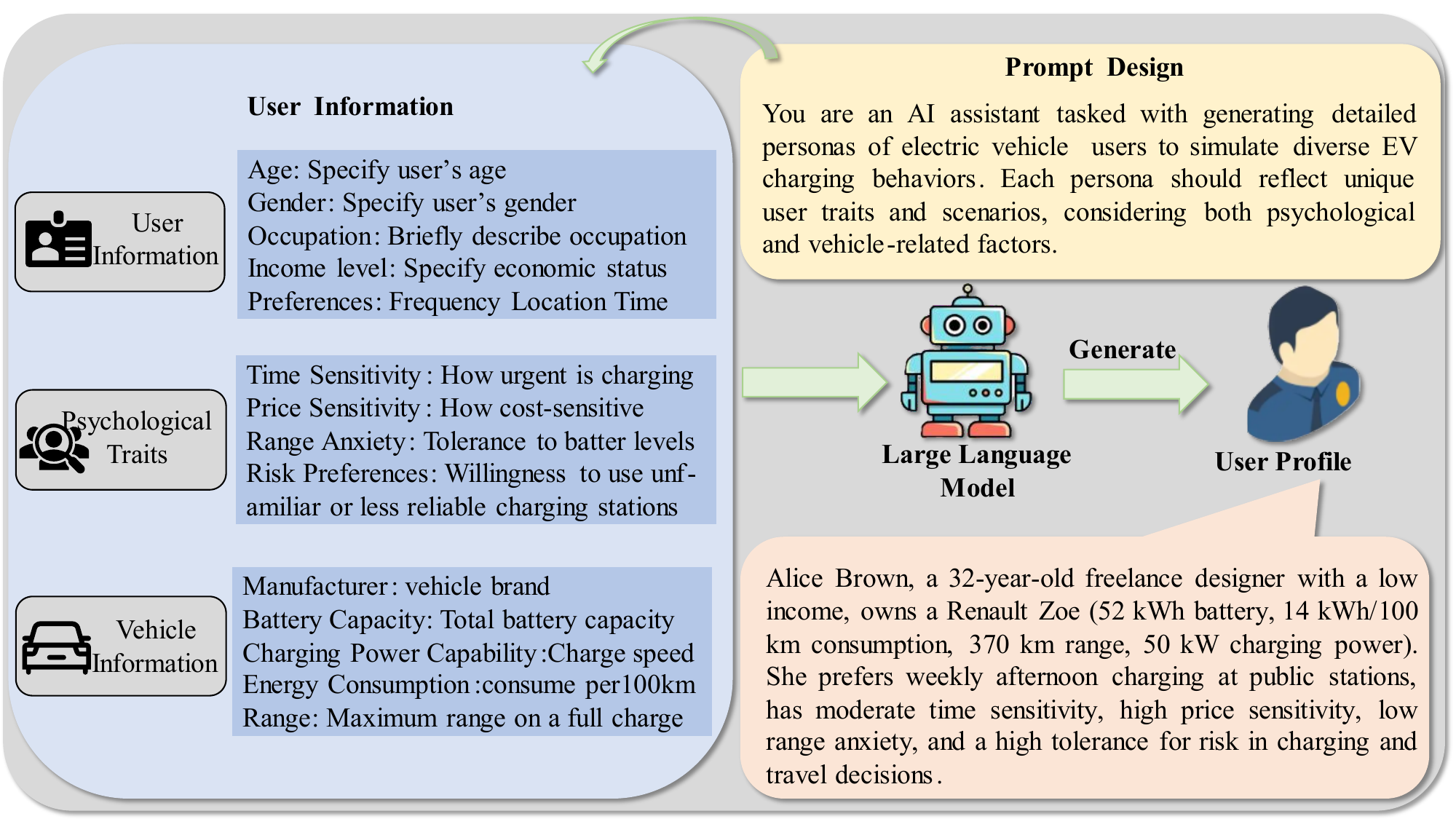}
    \caption{User profile generation with large language model}
    \label{fig:enter-label}
\end{figure}

User profiling encompasses three key dimensions: demographic characteristics, psychological traits, and charging behavior patterns.

\begin{itemize}
\item demographic characteristics include essential user attributes such as age, gender, occupation, income level, and place of residence. These attributes have a significant impact on users' charging behavior.
\item psychological traits significantly influence user decisions, encompassing environmental awareness, technology acceptance, and risk preferences. Users with strong environmental awareness are more inclined to adjust their charging behavior to support grid stability.
\item charging behavior patter define user preferences, encompassing charging frequency, time-of-day preferences, and charging location choices. Some users prioritize cost efficiency by charging during lower-priced periods, whereas others prioritize charging speed.
\end{itemize}

\subsection{Charging Behavior Decision Generation}
Simulating charging behavior decision-making is a core aspect of the user response model, designed to predict EV users' choices under varying electricity prices and environmental conditions.Utilizing user profile data and environmental state inputs, this study introduces an LLM-driven charging behavior decision-making method that dynamically models users' charging logic via a multi-agent system.
\paragraph{User Charging Behavior Response Model}

This paper establishes a user response model, as shown in \eqref{1}
, to describe the user's charging decision-making process and link charging behavior with environmental variables and dynamic electricity prices.The function $f_{\text{agent}_i}$  in the model is derived by the LLM through semantic reasoning on user profiles and environmental variables, accurately capturing user charging preferences and response behaviors under varying conditions.
\paragraph{Core Steps of User Decision Generation} 
The decision-making process for user charging behavior comprises three core steps:
\begin{itemize}
\item Charging intention assessment: Users decide whether to charge during the current time period based on the SoC.
\item Charging time selection:Users select an appropriate time period for charging based on dynamic electricity price signals and their time preferences.
\item Charging Amount Decision:Users determine the charging amount based on travel needs, dynamic electricity prices, and charging costs. 
\end{itemize}

Leveraging the semantic modeling capabilities of LLM, users' dynamic behavior decisions can be linked to the above logic, enabling more accurate simulation and prediction.

\paragraph{Multi-agent charging decision simulation}
To simulate charging behavior more accurately in multi-user scenarios, this study employs a multi-agent modeling approach, treating each user as an independent agent with decision-making logic driven by an LLM. 

\begin{itemize}
\item Each agent independently generates a charging behavior response based on personalized user profiles and environmental state inputs. 
\item The total charging demand from multiple agents is given by the following formula:
\begin{equation}
    P_{\text{total}}(t) = \sum_{i=1}^{N} P_{ri}(t)
    \label{eq}
\end{equation}
\end{itemize}

where $N$ is the total number of users, and $P_{\text{total}}(t)$ represents the total charging demand at time $t$.

Agent behaviors do not interfere with each other; however, batch simulations can evaluate the impact of different pricing strategies on overall charging load distribution.For example, dynamic pricing incentives during peak hours can significantly reduce users' simultaneous charging tendency, thereby balancing grid load. 

\paragraph{The dynamic characteristics of charging behavior decision-making}
This study's charging behavior decision-generation method fully accounts for dynamic environmental factors, including electricity price fluctuations, weather changes, and user travel demands. 
By incorporating real-time environmental variables into the LLM, user charging behavior simulation can be dynamically adjusted.
Price-sensitive users may delay charging when electricity prices increase.
In cold weather, users may charge more frequently at earlier times to mitigate range risks due to battery performance degradation. 
This dynamic response capability enhances the realism of charging behavior simulation and offers valuable insights for optimizing demand response strategies. 

\section{A two-level optimization model}

\subsection{Upper Level: Aggregator Pricing Model}

 In the electric vehicle demand response system, aggregators optimize their profits by adjusting retail electricity prices. The pricing model dynamically responds to grid load fluctuations and encourages users to charge during off-peak periods, thereby reducing peak load pressure on the power system. 
The aggregator's profit maximization problem is defined as follows: Over a given time interval, the aggregator adjusts the hourly retail electricity price $\lambda(t)$ to maximize profit, calculated as:
\begin{equation}
    \mathrm{Profit} = \mathrm{Total\ Revenue} - \mathrm{Total\ Cost}
    \label{eq}
\end{equation}

where total revenue is derived from user charging fees, while total cost consists of wholesale electricity purchase costs and fixed load consumption costs. Thus, the profit maximization problem is formulated as:
\begin{equation}
    \min_{\lambda(1),\dots,\lambda(H)} - \sum_{t=1}^{H} \left( \sum_{i=1}^{m} P_{ri}(t) + P f(t) \right) \cdot \left( \lambda(t) - \lambda_g(t) \right)
    \label{eq}
\end{equation}

Where $\lambda(t)$ represents the retail electricity price at time $t$, $\lambda_g(t)$ is the wholesale market electricity price at time $t$, $P_f(t)$ denotes the fixed load consumption at time $t$, and $P_{ri}(t)$ represents the actual electricity consumption of user $i$ at time $t$. The aggregator dynamically adjusts $\lambda(t)$ to encourage users to modify their charging behavior, thereby lowering wholesale electricity purchase costs and maximizing profits.

The consumer electricity consumption model enables demand response by dynamically adjusting the retail electricity price $\lambda(t)$.If the total electricity consumption $P_{ri}(t)$ has a nonlinear relationship with the retail price, the model uses multi-agent simulation to capture consumers' diverse response characteristics.Given a specific retail price, consumers decide whether to charge in the current time period based on their individual optimization strategies. 

In the optimization process, aggregators must adhere to upper and lower price constraints to maintain market stability and fairness. 

\begin{equation}
\text{s.t.} \quad \lambda_{\min} \leq \lambda(t) \leq \lambda_{\max}
\end{equation}

These constraints keep the price within the range \(\left[ \lambda_{\min}, \lambda_{\max} \right]\), preventing extreme fluctuations that may trigger excessive consumer responses or destabilize the market. 

\subsection{Lower-Level Model: User Behavior Modeling}
Modeling electric vehicle users' charging behavior is central to optimizing demand response systems, aiming to capture users' charging choices under varying electricity prices and incentives.The model represents users as intelligent agents whose charging behavior is influenced by personal attributes, vehicle characteristics, and environmental conditions.By incorporating multi-level user attributes, the model enhances the accuracy of predicting responses to electricity price fluctuations.

\begin{equation}
    P_{r_i}(t) = f_{\text{agent}_i}(\lambda(t))
    \label{eq:pri_function}
\end{equation}

$\text{Pri}(t)$ represents the total power consumption calculated by agent $m$ after receiving the electricity price $\lambda(t)$ based on its respective optimization strategy.
\begin{figure}[h]
    \centering
    \includegraphics[width=\linewidth]{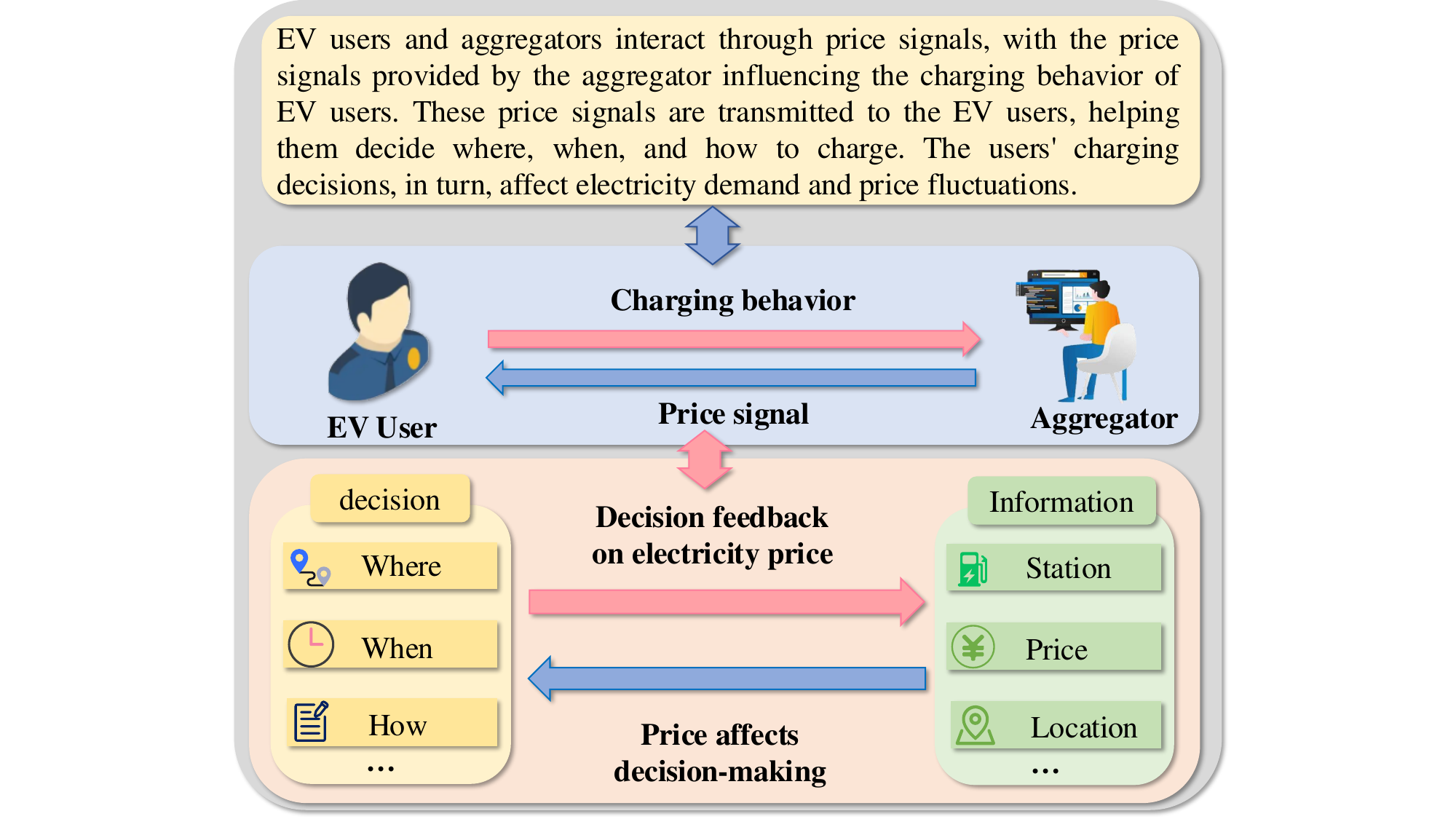}
    \caption{A TWO-LEVEL OPTIMIZATION MODEL}
    \label{fig:enter-label}
\end{figure}

\section{SIMULATION AND RESULTS}

In this study, we randomly generate various types of electric vehicle user and employ a large model to simulate their charging and discharging behaviors. In addition, the aggregator dynamically adjusts real-time electricity prices to influence user charging and discharging behaviors via demand response, with the aim of maximizing profits.The specific model parameters for this simulation are shown in Table 1.

\begin{table}[h]
\centering
\begin{tabular}{|c|c|c|c|}
\hline
definition & value  \\ \hline
DR time  & 06:00-11:00    \\ \hline
EV user  & 100   \\ \hline
retail price & \$0.09-\$0.22   \\ \hline
\end{tabular}
\caption{MODEL PARAMTERS}
\end{table}
The
specific simulation setup is as follows:

1) User Information Generation: LLM generates personal information, psychological characteristics, Vehicle information, and Charging demand of the users.

2) Charging Decision Generation: The LLM integrates user charging preferences with electricity price data to generate optimal charging decisions. 

3) The demand response simulation involves the aggregator adjusting real-time electricity prices to balance grid load and user demand, thereby maximizing profits.

\begin{figure}[h]
    \centering
    \includegraphics[width=\linewidth]{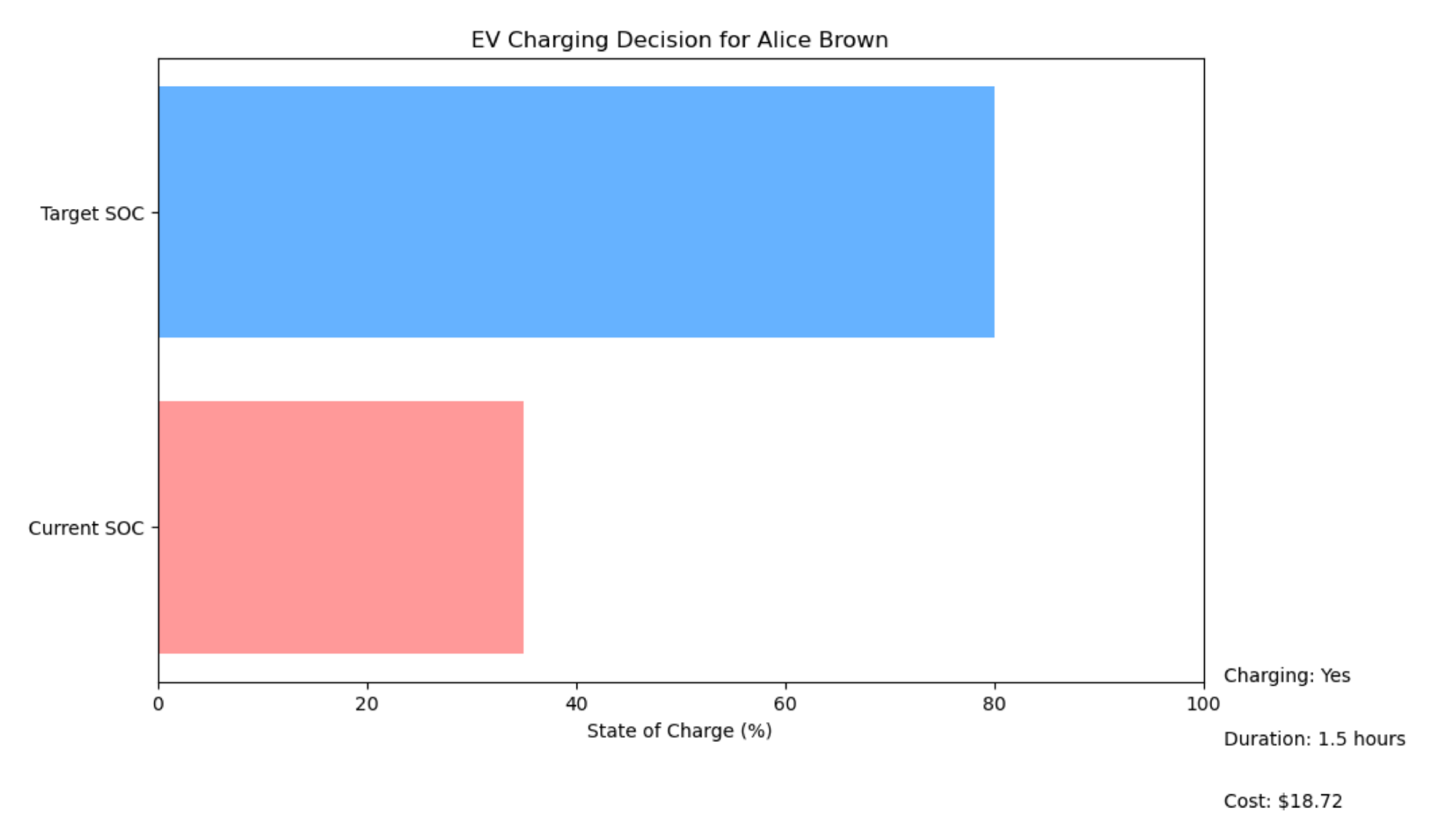}
    \caption{charging decision generation}
    \label{fig:enter-label}
\end{figure}

As shown in Figure 3, the method proposed in this study effectively generates user-specific charging decisions.For example, for Alice, whose vehicle’s current State of Charge (SoC) is 30\% and whose charging demand is to reach 80\% to meet her next travel needs, the large language model, integrating the current electricity price and her travel requirements, determines that charging is required. The charging duration is set to 1.5 hours, with a cost of 18.72. 

\begin{figure}[h]
    \centering
    \includegraphics[width=\linewidth]{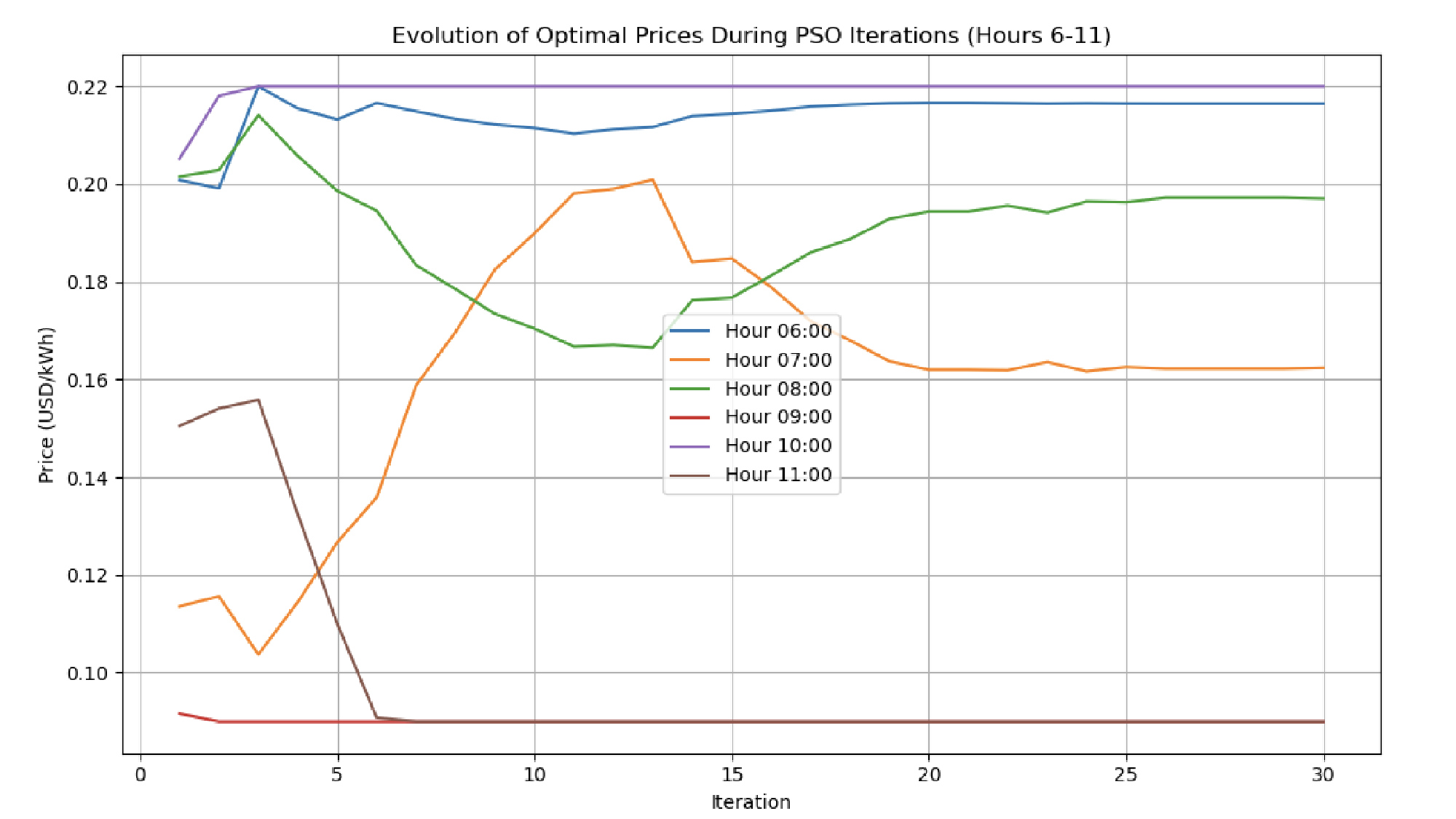}
    \caption{optimize electricity price}
    \label{fig:enter-label}
\end{figure}

As shown in the simulation results in Figure 4, Particle Swarm Optimization (PSO) helps the aggregator optimize electricity prices.Initially, the aggregator may explore multiple possible price ranges. As iterations progress, the aggregator adjusts electricity prices based on user charging demands, ultimately ensuring profitability.

\section{CONCLUSION}

This paper proposes a novel two-layer demand response framework based on Large Language Models (LLM), designed to optimize collaboration between Electric Vehicle (EV) users and aggregators, achieving grid load balancing and efficient energy utilization.
The upper-layer model focuses on maximizing the aggregator's profit through dynamic electricity price adjustments, while the lower-layer model uses LLM to simulate users' charging behavior under different price and environmental conditions.Key innovations include:

1)LLM-driven User Behavior Simulation Framework: By integrating demographic features, psychological preferences, and charging behavior patterns, the framework accurately captures user heterogeneity, enhancing the personalized adaptation of demand response strategies. 

2)Dialog-Based Decision Support Method: Leveraging the semantic reasoning capabilities of LLM, dynamic charging recommendations are provided, balancing cost efficiency with comfort preferences. 

3)Two-layer Optimization Mechanism: Coordinating the aggregator's profit objectives with user charging demands, alleviating grid peak load pressures, and ensuring a dynamic supply-demand balance.

\vspace{10pt}

\thispagestyle{plain} 
\end{document}